\documentclass[11pt,a4paper]{article}

\usepackage[margin=1in]{geometry}
\usepackage{amsmath,amssymb,amsfonts}
\usepackage{booktabs}
\usepackage{enumitem}
\usepackage{hyperref}
\usepackage[numbers]{natbib}
\usepackage{xcolor}
\usepackage{float}
\usepackage{authblk}
\usepackage{titlesec}
\usepackage{array}
\usepackage{ragged2e}

\newcolumntype{L}[1]{>{\RaggedRight\arraybackslash}p{#1}}

\hypersetup{
    colorlinks=true,
    linkcolor=blue!60!black,
    citecolor=blue!60!black,
    urlcolor=blue!60!black
}

\titlespacing*{\section}{0pt}{1.5ex plus 0.5ex minus .2ex}{1ex plus .2ex}
\titlespacing*{\subsection}{0pt}{1.2ex plus 0.3ex minus .2ex}{0.8ex plus .2ex}
\titlespacing*{\subsubsection}{0pt}{1ex plus 0.2ex minus .2ex}{0.6ex plus .1ex}

\title{Votiverse: A Configurable Governance Platform\\for Democratic Decision-Making}

\author[1]{Diego Macrini}
\affil[1]{Proximify Inc., Ottawa, Canada\\ \texttt{dmac@cs.toronto.edu}}
\date{March 2026}

\begin{document}

\maketitle

\begin{abstract}
Democracy is not a single mechanism. It is a space of possible configurations---a spectrum stretching from pure direct participation to full delegation of authority. The systems we live under today occupy a narrow band of that spectrum, chosen centuries ago under constraints that no longer apply, and rarely questioned since.

Votiverse is a platform for exploring the rest of that space. It provides organizations, communities, and institutions of any size with a configurable governance engine. Participants can vote directly, delegate their vote to trusted individuals by topic, or operate under any hybrid arrangement their group defines. Delegations are revocable, topic-specific, and transitive. A direct vote always overrides a delegation. In this model, traditional representative democracy is not the norm---it is an edge case: the configuration you get when delegation is forced, universal, non-specific, and irrevocable for a fixed term.

Votiverse introduces two structural innovations. First, a \emph{governance awareness layer}---a built-in intelligence system that monitors the delegation network and delivers contextual, progressive-disclosure reporting to participants at the point of decision. Concentration patterns, chain anomalies, and delegate track records are not buried in dashboards; they surface when and where a participant needs them. Second, a \emph{prediction-tracking accountability layer}. Proposals carry falsifiable predictions. Outcomes are recorded. Over time, the platform builds a collective memory of what was decided, what was promised, and what actually happened. Together, these layers transform voting from a momentary act into an ongoing process of collective learning.

This paper formalizes the governance model, situates it within existing work on liquid democracy and participatory decision-making, addresses known failure modes, and describes the architecture of the platform. The core platform has been implemented and released as open-source software.
\end{abstract}

\section{The Problem}

Every group that must make collective decisions faces two interrelated problems.

\subsection{The Participation Problem}

\textbf{Direct democracy} asks every participant to engage with every decision. This works for small groups and important issues, but it does not scale. As the number of decisions grows, participation collapses under the weight of its own demands. People disengage---not because they don't care, but because there are not enough hours in the day to form an informed opinion on everything.

\textbf{Representative democracy} solves the bandwidth problem by bundling all decisions into a single delegation. You choose one person (or party) to decide on your behalf, on all matters, for a fixed period. This scales well but discards nearly all information about what voters actually want. A citizen who cares deeply about environmental policy, disagrees with their representative on education, and has no opinion on trade must accept the entire package or abandon the representative entirely. The resolution is far too low.

These two systems are typically presented as the only options. They are not. They are extremes of a continuum---and the interesting territory lies between them.

\subsection{The Information Problem}

There is a second failure, less discussed but arguably more damaging: \textbf{existing democratic systems do not close the loop between decisions and outcomes.}

Politicians and parties make proposals. Those proposals contain implicit or explicit predictions---``this policy will create jobs,'' ``this spending will reduce crime,'' ``this reform will lower costs.'' Voters choose between these competing predictions. The winning proposal is implemented. And then\ldots\ nothing. There is no structured record connecting the original promise to the actual outcome. No systematic way to ask: was the proposal implemented? Did it perform as predicted? Who predicted accurately, and who didn't?

This information exists, scattered across government reports, academic studies, and journalistic investigations. But it is not connected to the original proposals, not surfaced at the moment of the next decision, and not accessible to the ordinary voter. The result is that political discourse collapses into narrative---who is likeable, who is trustworthy, which team is winning---rather than evidence.

\subsection{What Is Needed}

What is needed is not just a better voting mechanism. It is a better \textbf{information infrastructure for collective decision-making}---one that records proposals, predictions, and outcomes; connects them over time; surfaces them at the point of decision; and identifies, structurally and without bias, who predicts well and who doesn't.

The voting mechanism matters---selective engagement, topic-specific delegation, revocability. But the information layer may matter more. Even in a system where most participants delegate passively, the quality of governance improves if the information environment is honest, structured, and persistent.

That mechanism exists in theory. It has been called \emph{liquid democracy}, \emph{delegative voting}, or \emph{proxy voting}. Votiverse is an attempt to combine it with the missing information infrastructure and make both practical, configurable, and accountable at every scale---from a neighborhood sports club to a national government.

\section{Principles}

The design of Votiverse is guided by seven principles.

\textbf{Participation without burden.} No one should be forced to have an opinion on everything, and no one should be prevented from having an opinion on anything. The system must make it easy to engage selectively.

\textbf{Sensing as participation.} Governance should not only ask people to decide---it should ask them to observe. Every participant, regardless of expertise or engagement level, is a sensor embedded in their local reality. Structured, accountable surveying of lived experience is a form of participation that is as valuable as voting and far more inclusive.

\textbf{Expertise without permanent power.} People should be able to channel their vote through those they trust on specific topics. But that trust must be revocable, granular, and never locked into a fixed term. A delegate's influence should last exactly as long as it is deserved.

\textbf{Accountability through prediction.} A decision is only as good as the reasoning behind it. Proposals should include falsifiable predictions about their expected outcomes. When reality diverges from prediction, the record should be clear and accessible.

\textbf{Active awareness over passive transparency.} It is not enough to make data available. The system must actively inform participants about what matters, when it matters. Delegation chains, concentration patterns, delegate track records, and prediction outcomes must surface at the point of decision---contextually, progressively, and without overwhelming the voter.

\textbf{Configurability over prescription.} Different groups have different needs. A soccer parents' committee, a cooperative, and a nation-state require different governance configurations. Votiverse provides the primitives; organizations compose them to suit their context.

\textbf{Scale independence.} The same fundamental mechanisms---voting, delegation, structured proposals, prediction tracking, participant surveys---should work whether the group has twelve members or twelve million.

\section{The Scale Problem}

\subsection{What We Know About Humans and Group Decisions}

Humans evolved to make collective decisions in small groups---roughly 30 to 150 people. At that scale, participants know each other, see the consequences of decisions directly, and feel their individual contribution matters. Governance is personal, concrete, and self-correcting.

As groups grow beyond this range, something breaks. The consequences of decisions become abstract. The individual's influence becomes statistically invisible. The feedback loops that connect a decision to its outcome become slow, diffuse, and hard to attribute. The rational response---genuinely rational---is to disengage and let someone else handle it. Political scientists call this \emph{rational ignorance}: the cost of becoming informed exceeds the expected benefit of casting a slightly better-informed vote.

\subsection{The Propaganda Vulnerability}

Disengagement creates a vacuum, and that vacuum is filled by whoever controls the narrative. When people process governance through tribal heuristics---party identity, charismatic figures, social media momentum---rather than through deliberation, they become susceptible to manipulation. This is true in every existing democratic system, and a more sophisticated voting platform does not inherently fix it.

A delegation system could even make it worse. ``I delegated to an expert'' is a more comfortable form of disengagement than ``I voted for whichever party my family always votes for,'' but the cognitive posture is the same: someone else is handling it.

\subsection{What Votiverse Can and Cannot Do}

Votiverse does not claim to solve the fundamental disconnect between human cognitive capacity and the scale of modern governance. No platform can. The root cause is not technological---it is the mismatch between the scale at which human cognition works well and the scale at which modern governance operates.

What Votiverse can do is create better conditions at every scale:

\textbf{At small scale}---clubs, cooperatives, teams, local associations---the platform provides structure and accountability for decisions that people already care about. Participants are cognitively engaged because they know each other, see the consequences, and feel their stake.

\textbf{At medium scale}---organizations, municipalities, institutions---delegation and prediction tracking provide real value. The trust network makes expertise accessible without creating a permanent political class. The prediction record grounds decisions in evidence. The awareness layer keeps power dynamics visible.

\textbf{At national scale}---the hardest case---Votiverse may primarily serve as infrastructure that makes existing power dynamics \emph{visible} rather than hidden. Even if most participants delegate passively, the system ensures that the concentration of power is measured, reported, and available for the moments when enough people do re-engage.

\subsection{Sensing and Deciding: Two Different Asks}

The scale problem becomes less daunting once we recognize that governance asks participants to do two fundamentally different things---and only one of them is hard.

\textbf{Deciding} asks: ``what should we do?'' This requires understanding proposals, evaluating trade-offs, predicting consequences, and exercising judgment. It is cognitively demanding. Most people are not good at it for most topics, and the rational response at scale is to disengage or delegate.

\textbf{Sensing} asks: ``what is happening?'' Is your neighborhood safer? Has traffic improved? Are prices higher? Is the school better than last year? This is observation---raw, local, personal. It requires no expertise, no policy knowledge, no deliberation. It requires only that you report what you experience in your daily life.

This distinction is fundamental to Votiverse's design. The platform separates sensing from deciding. Votes and delegations handle deciding. Participant surveys handle sensing. And critically, sensing feeds deciding: survey data---community observations, trend lines, ground-level feedback---flows into the awareness layer and is surfaced at the point of decision when new proposals are introduced.

\subsection{Working With the Grain: Topic Communities}

Topic-specific delegation effectively creates \textbf{virtual communities of interest} within any large-scale deployment. The person who cares about education policy is part of an implicit group of people who also care about education---delegates, delegators, and direct voters on education issues. That group is small enough to be cognitively manageable, even within a municipality of fifty thousand.

The awareness layer can lean into this. Instead of asking a participant to comprehend the entire governance system---every delegation, every issue, every outcome---it can surface the dynamics of their topic communities.

\subsection{Open Questions}

Several questions remain unresolved:

\textbf{Does delegation reduce deliberation?} If delegation is too easy, do participants stop reading the booklet entirely? Does the system converge toward a small number of super-delegates who function as de facto representatives?

\textbf{Does prediction tracking change behavior?} The hypothesis is that visible, falsifiable predictions create a feedback loop that makes governance quality tangible. This is plausible but unproven at scale.

\textbf{Is national-scale liquid democracy meaningfully better than representative democracy?} The honest answer is that we do not know, and the project does not depend on the answer being yes. Votiverse is valuable if it improves decision-making at \emph{any} scale.

\textbf{Survey fatigue.} If surveys are the primary evidence base, they must recur regularly. What is the optimal cadence before participants disengage?

\textbf{The limits of self-generated evidence.} Can a community accurately assess outcomes that affect it? Cognitive biases may systematically distort participant observations. The longitudinal structure of surveys mitigates this but does not eliminate it.

These questions are not reasons to abandon the project. They are reasons to be honest about what it is: an attempt to expand the space of possible governance configurations, tested first at the scales where improvement is most likely, and extended carefully toward larger scales as evidence accumulates.

\section{The Governance Configuration Space}

\subsection{Primitives}

Every participant can cast a \textbf{direct vote} on any issue. This is the irreducible primitive---the act of governance itself. When delegation is available, three structural properties hold unconditionally and are not configurable: \emph{revocability} (any delegation can be withdrawn at any time before the vote closes), \emph{sovereignty} (a direct vote always overrides any delegation), and \emph{scope} (delegations can be scoped to a topic or a single issue, and this granularity is always available). These are not design parameters---they are rights that the platform guarantees.

What \emph{is} configurable is how delegation works. The delegation model is defined by two independent boolean axes.

\subsection{The Two Axes of Delegation}

\textbf{Candidacy.} Is there a formal system for declaring ``I am willing to represent others''? When enabled, participants can publish structured profiles, state their positions, and appear in the delegate discovery interface. This is an accountability mechanism: people accumulating delegated voting power have explicitly opted in and made their positions visible. When disabled, delegation (if it exists) is informal---you delegate to someone you know personally, without institutional infrastructure for discovering or evaluating potential delegates.

\textbf{Transferability.} Can delegated voting power flow through chains? When enabled, if Alice delegates to Bob and Bob delegates to Carol, then Carol carries Alice's weight. Voting power is transferable---it flows through the delegation graph. This is the core mechanism of liquid democracy. When disabled, delegated power cannot be transferred beyond one hop.

These two axes produce four governance families:

\begin{table}[H]
\centering
\begin{tabular}{@{}l|L{5.2cm}|L{5.2cm}@{}}
\toprule
 & \textbf{No transfers} & \textbf{Transfers} \\
\midrule
\textbf{No candidates} &
\emph{Direct democracy.} Everyone votes on everything. No delegation mechanism exists. &
\emph{Informal liquid.} Anyone can delegate to anyone. Chains flow freely. No formal candidate system. Delegators rely on personal knowledge to choose delegates. \\
\midrule
\textbf{Candidates} &
\emph{Representative.} Appoint a declared candidate as your proxy. They vote for you but cannot pass your vote further. Classic proxy voting. &
\emph{Liquid delegation.} Candidates exist for discoverability and accountability. Anyone can still delegate to anyone. Chains are transitive. \\
\bottomrule
\end{tabular}
\end{table}

Delegation exists when either axis is enabled. When both are off, the system is pure direct democracy. This 2$\times$2 is not a simplification of a richer space---it \emph{is} the space. The familiar governance models map naturally to this grid. Traditional representative democracy is the bottom-left: candidates exist, but your vote stops with them. Liquid democracy as theorized by Tullock~\cite{tullock1967}, Miller~\cite{miller1969}, and Ford~\cite{ford2002} is the top-right: delegation is transitive and revocable, but there is no formal candidacy system. The bottom-right adds the candidacy profiles that make transitive delegation possible where delegators do not personally know their delegates.

\subsection{Delegate Candidacies as Proposals}

When candidacy is enabled, a delegate candidacy is not merely a profile---it is a \textbf{proposal} submitted for community evaluation. A candidate publishes structured claims: qualifications, stated positions, relevant experience. These claims share the same scrutiny infrastructure as policy proposals: community notes can annotate them, version history is preserved, and the profile is immutable once active (updates create new versions; old versions persist). This unification means the platform needs a single accountability infrastructure for both kinds of claims---policy proposals and leadership candidacies---rather than two parallel systems.

The distinction between formal and informal delegation matters here. When a participant delegates to someone they know personally (``I trust Maria on education topics''), that is an exercise of a right, not a claim submitted for scrutiny. But when someone publishes a candidacy seeking delegations from strangers, they are making a public claim that warrants public evaluation. Candidacies also enable \emph{opt-in vote transparency}: a candidate can choose to let delegators see how they voted on every issue within the delegation scope. This is a per-candidate signal of trustworthiness, not an organization-wide setting.

\subsection{Beyond Delegation: Ballot, Features, and Timeline}

The delegation axes determine how voting power flows. Orthogonal to this, three further dimensions complete the governance configuration: \emph{ballot rules} (secret vs.\ public ballot, sealed vs.\ live results, vote mutability, quorum, voting method), \emph{feature toggles} (community notes, prediction tracking, surveys), and \emph{timeline} (deliberation, curation, and voting periods). Together with the two delegation axes, these produce the full parameter space---small enough for any group creator to configure, expressive enough for meaningful governance research.

\subsection{Named Presets}

Presets are named points in the parameter space, each representing a distinct governance philosophy---not a parameter tweak. The platform currently provides six presets: \emph{Direct Democracy} (no delegation, minimal infrastructure), \emph{Swiss Votation} (direct democracy with structured booklets, community notes, and predictions), \emph{Informal Liquid} (informal liquid delegation without candidates, public ballots, short timelines), \emph{Representative} (declared candidates, non-transitive proxy voting, high quorum), \emph{Liquid Delegation} (the full composition: candidates with transitive delegation, structured booklets, community notes, predictions, surveys, and awareness), and \emph{Civic Participatory} (liquid delegation at municipal scale with longer timelines). Presets are starting points; organizations customize from there.

\subsection{The Continuum}

The key insight is that representative democracy and direct democracy are not opposites. They are two of four quadrants in a 2$\times$2 defined by candidacy and transferability. Most real governance needs fall in the quadrants that existing systems leave unexplored---particularly the bottom-right, where delegation is transitive, revocable, and supported by accountability infrastructure. Votiverse makes these quadrants accessible.

\section{How Voting and Delegation Work}

This section describes how Votiverse resolves votes in the presence of delegations. A full formal specification is provided in Appendix~\ref{app:formal}.

\subsection{The Basic Mechanism}

Every participant in a Votiverse instance can do one of three things for any given issue: vote directly, delegate, or abstain.

A \textbf{delegation} is an instruction: ``If I don't vote on this issue, let this person's vote count for me too.'' It has three components: the person delegating, the person receiving the delegation, and the \textbf{scope}---which topics or issues the delegation covers.

Delegations can be broad (``delegate all Finance topics to Alice'') or narrow (``delegate this specific budget proposal to Alice''). Delegations operate at three levels of scope: \textbf{global} (all issues), \textbf{topic} (issues classified under a specific topic or its descendants), and \textbf{issue} (a single specific issue). When multiple delegations from the same person are active for the same issue, the most specific one wins: issue-scoped overrides topic-scoped, child-topic overrides parent-topic, and any topic-scoped delegation overrides a global one. All three scope types are transitive. Issue-scoped delegations are ephemeral---they apply to one decision and do not persist. A participant can hold at most one delegation per scope: delegating Education to Carol after having delegated Education to Bob simply replaces the earlier delegation.

\subsection{The Override Rule}

The most important rule in the system: \textbf{a direct vote always wins.}

If you have delegated Climate Policy to Maria, but a particular climate proposal matters to you, you simply vote. Your vote counts as your own, with weight one, and Maria's vote does not include yours. You do not need to revoke the delegation---your direct participation automatically overrides it for that issue.

This is the ``backup'' semantics at the heart of the system. A delegate is not your representative in the traditional sense. They are your fallback---the person who votes on your behalf only if you choose not to.

\subsection{Transitive Delegation}

Delegations can flow through chains. Suppose Alex delegates Finance to Beth, and Beth delegates Finance to Carlos. If none of them vote directly, Carlos casts one vote that carries the weight of all three: his own, Beth's, and Alex's. This is \textbf{transitive delegation}: trust flows through the network.

There is no limit on chain length. Rather than imposing artificial depth caps---which can silently lose votes when downstream delegations change---Votiverse relies on the awareness layer's chain-resolution display to keep participants informed about who ultimately casts their vote.

\subsection{How Overrides Interact with Chains}

Overrides break chains cleanly. Consider the chain Alex $\to$ Beth $\to$ Carlos on a Finance issue:

If Beth votes directly, Beth's vote counts with weight two (her own plus Alex's delegation). Carlos carries weight one. If Alex votes directly, Alex's vote counts with weight one, and Carlos carries weight two (his own plus Beth's). If all three vote directly, each carries weight one---pure direct democracy.

The principle is consistent: a direct vote severs the chain at that point.

\subsection{Cycles}

Transitive delegation can create cycles. Votiverse resolves cycles simply: if participants form a cycle and none of them vote directly, all participants in the cycle are treated as abstaining. If any participant in the cycle votes directly, that breaks the cycle at their position, and the rest of the chain resolves normally.

\subsection{Guarantees}

The model guarantees four properties that hold regardless of how delegations are configured:

\textbf{Sovereignty.} Every participant can always vote directly. No delegation can prevent someone from casting their own vote.

\textbf{One person, one vote.} Every participant contributes exactly one unit of voting weight to the final tally. Delegation moves weight through the network---it never creates or destroys it.

\textbf{Monotonicity.} Voting directly never makes you worse off. Casting a vote can only increase your influence on the outcome compared to delegating.

\textbf{Revocability.} Any delegation can be withdrawn at any time before the vote closes. The system recomputes weights immediately.

\section{Voting Events and the Digital Booklet}

\subsection{Voting Events}

A voting event is a structured period during which one or more issues are put to a vote. It is the operational unit of governance in Votiverse, analogous to a Swiss \emph{votation day}. During deliberation, individual issues may be \textbf{cancelled}---for example, to correct a topic misclassification or withdraw an issue that is no longer relevant. Cancelled issues cannot receive votes, and any active proposals associated with them are automatically withdrawn. Cancellation is permanent and recorded; a cancelled issue cannot be reopened.

\subsection{The Digital Voting Booklet}

Inspired by the Swiss Federal Council's practice of mailing a physical booklet to every citizen before a vote, Votiverse provides a \textbf{digital voting booklet} for each issue. When enabled, the booklet is a structural requirement: no issue can proceed to vote without a complete booklet. Organizations that do not need structured deliberation can disable the booklet entirely.

For each issue, the booklet contains: (1) an official description---a neutral summary prepared by the event administrators; (2) proposal text---the specific proposal being voted on; (3) supporting arguments submitted by the proponents; (4) opposing arguments submitted by opponents or administrators; (5) falsifiable predictions about expected outcomes; and (6) current state of affairs---relevant metrics and context at the time of voting.

\subsection{Roles}

Three roles govern the lifecycle of a voting event. \textbf{Administrators} create voting events, define issues, manage timelines, and enforce formatting rules---though in community governance they are typically volunteer participants, not neutral third parties, so the platform does not rely on them for content curation or claim verification. \textbf{Proponents} submit proposals, provide supporting arguments and predictions. \textbf{Members} read the booklet, participate in deliberation, and vote.

\section{Prediction Tracking and Accountability}

\subsection{The Accountability Gap}

Traditional democratic systems record what was decided but not what was expected. A proposal wins, a policy is implemented, and years later, no one systematically compares the outcome to the promises that secured the votes. This is the accountability gap.

\subsection{Proposals as Models}

There is a foundational principle from probabilistic reasoning that applies directly to governance: \textbf{the model that predicts best is the one you should trust.}

In Bayesian model selection, competing models each tell a story about how the world works. The test of a model's quality is not how well it explains the past but how well it predicts data it has not yet seen. Governance proposals are models. Each proposal is an implicit claim about how the world works: ``if we adopt this policy, these outcomes will follow.''

The current political system has no holdout set. It never systematically checks predictions against outcomes. It evaluates proposals entirely on narrative fit. This is the equivalent of selecting a statistical model based on its training error alone, with no cross-validation. It is a system optimized for overfitting.

Prediction tracking is the regularizer. It introduces a cost for overconfident, unfalsifiable, or inaccurate claims by checking them against outcomes. Over time, it builds the equivalent of a cross-validation record for governance.

\subsection{Predictions as First-Class Objects}

In Votiverse, predictions are a core feature of the platform, configurable per organization. When enabled, every proposal is encouraged---or, at the organization's discretion, required---to include one or more falsifiable predictions.

A prediction is a structured claim containing: a measurable variable, a direction and magnitude, a timeframe, and an optional methodology.

\subsection{Evaluation}

At the end of the prediction's timeframe, an evaluation is triggered. The primary evaluation mechanism is community-driven: participants assess the outcome through structured surveys (Section~9). AI-assisted monitoring of public data sources and official statistics provide supplementary evidence where available. Predictions are scored on a structured scale from ``clearly met'' through ``partially met'' to ``clearly not met'' and ``unfalsifiable.''

\section{Community Notes}

Community notes are participant-submitted annotations attached to proposals, arguments, predictions, or any other element in the voting booklet. They are inspired by the community notes feature on social media platforms, adapted for a governance context.

A community note provides context, correction, or additional evidence. It is not a comment or opinion---it is a factual annotation. Notes are evaluated by other participants: endorsed as helpful or disputed as misleading. The evaluation mechanism uses a bridging criterion: a note earns prominent visibility only when it is rated helpful by participants across different viewpoints.

This bridging requirement is critical. It filters out partisan cheerleading and surfaces information that is genuinely informative---claims endorsed by both supporters and opponents of the proposal.

Because delegate candidacies and policy proposals share the same scrutiny infrastructure (Section~4), community notes apply uniformly to both---as well as to survey results and to other community notes. Notes on notes enable dispute resolution without administrative intervention. This is essential because community governance typically has no neutral administrators; the ``administrator'' is a volunteer participant, not a third party. The system must assume that verification is distributed among participants rather than delegated to an operator.

\section{Participant Surveys}

\subsection{Sensing as a Primitive}

Surveys are a fundamental governance primitive in Votiverse. They are the mechanism through which the platform captures participant observations---the sensing layer that complements the deciding layer of votes and delegations.

\subsection{Surveys Are Not Votes}

The distinction is fundamental. A \textbf{vote} expresses a decision. It can be delegated, because someone you trust can exercise judgment on your behalf. A \textbf{survey} expresses an observation. It cannot be delegated, because the entire point is \emph{your} experience from \emph{your} position. Survey responses are therefore \textbf{non-transferable}.

\subsection{Accountable and Open}

Votiverse surveys are structurally different from conventional polling: open participation (every eligible participant can respond), visible questions (publicly inspectable framing), transparent results (raw data available to all), and accountable provenance (who proposed what is recorded).

\subsection{Surveys as Ground Truth for Predictions}

Participant surveys are the \textbf{primary evidence base} for prediction verification. AI-assisted monitoring of external data sources is a useful supplement for publicly measurable variables, but for the internal, qualitative outcomes that matter most in community governance---whether park maintenance improved, whether communication got better, whether the new scheduling system works---the community itself is the oracle. External data often does not exist for these questions.

This makes surveys a structural requirement for the accountability system, not a secondary engagement feature. The evidential loop is: proposals make predictions $\to$ surveys capture observations $\to$ community notes link surveys to predictions $\to$ participants see whether predictions held $\to$ future proposals carry more or less credibility. The system generates, stores, evaluates, and surfaces its own evidence without requiring external data sources or human fact-checkers.

\subsection{Surveys as Trend Lines}

Because surveys recur on a predictable schedule, they produce longitudinal trends. A recurring survey yields a time series of community observation that runs in parallel with the governance timeline. This transforms surveys from snapshots into a monitoring system.

\section{The Governance Awareness Layer}

\subsection{From Passive Transparency to Active Awareness}

Votiverse includes a \textbf{governance awareness layer}: a built-in system that continuously monitors the state of the delegation network, the history of decisions and outcomes, and the behavior of delegates---and delivers relevant findings to participants contextually, at the moment they are making decisions. The analogy is not a transparency dashboard. It is a built-in newsroom that reports on the health of the governance system itself.

\subsection{Design Principles}

The awareness layer is governed by four principles: \emph{contextual delivery} (information surfaces at the point of decision), \emph{progressive disclosure} (summary first, detail on demand), \emph{personal relevance} (personalized notifications based on what affects your vote), and \emph{signal over completeness} (anomalies and high-signal findings prioritized over routine data).

\subsection{What the Awareness Layer Surfaces}

The layer draws on three sources---the delegation network, the decision history, and the prediction record---and surfaces findings including: delegation chain resolution, concentration alerts, delegation harvesting detection, delegate track records, engagement prompts, and historical context.

\subsection{Personal Voting History}

The awareness layer maintains a personal voting history---a retrospective record of every decision a participant was involved in. Over time, this accumulates into a voting footprint that enables self-correction: a participant can see whether their delegates' backed proposals actually delivered on their predictions.

\section{Risks and Mitigations}

\subsection{Power Concentration}

\textbf{Risk.} Transitive delegation can create super-delegates who accumulate enormous voting weight. \textbf{Mitigation.} The governance awareness layer provides real-time concentration metrics, personalized alerts, delegation harvesting detection, and full chain-resolution visibility. Delegations are revocable at any time.

\subsection{Delegation Overuse}

\textbf{Risk.} People may delegate too readily, deferring to perceived experts even when they would make better decisions themselves. \textbf{Mitigation.} Delegation is not the default. The platform encourages direct voting by presenting the booklet and making direct participation easy. The awareness layer's engagement prompts provide contextual nudges.

\subsection{Strategic Behavior}

\textbf{Risk.} Delegates may act strategically---accumulating delegations on a popular stance and then switching positions. \textbf{Mitigation.} The awareness layer makes delegate behavior visible through track records, voting history, and prediction accuracy.

\subsection{Long Delegation Chains}

\textbf{Risk.} Unlimited transitivity can produce chains where source participants have no idea who ultimately casts their vote. \textbf{Mitigation.} The awareness layer's chain-resolution display ensures participants always know who ultimately votes on their behalf. Rather than imposing artificial depth caps---which can silently lose votes when downstream delegations change---the platform relies on transparency and contextual warnings.

\subsection{Digital Divide and Accessibility}

\textbf{Risk.} A digital governance platform inherently excludes people without reliable internet access or digital literacy. \textbf{Mitigation.} Offline booklets, assisted voting, and progressive deployment starting where digital access is already the norm.

\section{Identity, Trust, and Scale}

Votiverse supports a spectrum of identity models: invitation-based (small groups), organizational authentication (medium groups), verified identity (large civic deployments), and cryptographic identity (decentralized deployments). At every scale, the platform must resist Sybil attacks---the creation of fake participants to multiply voting power---with defenses that vary by identity model.

Delegation graphs contain sensitive information. The platform balances transparency (necessary for accountability) with privacy (necessary for free participation without coercion), with configurable visibility per organization.

\section{Platform Architecture}

Votiverse is not a single application. It is a \textbf{configurable governance engine} that organizations instantiate with their own settings. Each instance defines the set of governance primitives in use, the topic taxonomy, the identity model, the voting rules, prediction tracking configuration, delegation constraints, booklet requirements, and the visibility model for delegations and votes.

The architecture is designed for real-time delegation graph computation, configurable access control, audit logging of all governance actions, and data portability.

\subsection{Immutability Guarantees}

The self-maintenance mechanisms described in this paper---community notes, prediction tracking, delegate candidacies, surveys---all depend on a single precondition: \textbf{the record cannot be altered after the fact.} If proposals can be silently edited after community notes reference them, the notes become meaningless. If survey responses can be modified, the evidence base is unreliable. If delegate profiles can change without versioning, accountability is theater.

Votiverse distinguishes between what must never change and what may. The event store is append-only: events are never modified or deleted. Proposals are immutable once submitted for voting. Delegate profile versions are immutable once active (updates create new versions; old versions persist). Survey responses and community notes are immutable once submitted. Conversely, votes may be changed during an open voting period (configurable), and delegations may be created or revoked at any time---trust is inherently mutable.

\subsection{The Role of AI---and Its Limits}

AI can play a valuable role in the information layer---particularly in gathering evidence that confirms or contradicts predictions. However, \textbf{AI systems are not neutral.} Votiverse's approach follows the platform's own principles: transparency (which AI system is used), auditability (sources inspectable), replaceability (organizations can switch providers), separation of roles (AI gathers information, humans judge), and ensemble verification (multiple AI providers operating in parallel).

\subsection{Platform Integrity: Blockchain and Oracles}

Blockchain technology provides an optional \textbf{integrity layer}---not as the foundation of the platform, but as a tamper-evident seal on critical integrity artifacts: vote tallies, prediction commitments, outcome recordings, survey results, and delegation graph snapshots. The combination of multiple oracle sources---AI, official data providers, surveys, and community challenge---provides resilience through mutual verification.

\subsection{Implementation and Availability}

The governance engine and platform logic described in this paper have been implemented and are available as open-source software under AGPL-3.0 licensing at \url{https://github.com/votiverse/votiverse}. Public access is provided through web and mobile applications (iOS and Android) at \url{https://votiverse.app}. The public service is free to use; its infrastructure is funded by organizational sponsors, currently Proximify Inc. The deployed service uses the open-source core together with a thin private layer for application packaging and operational infrastructure.

\section{Deployment Strategy}

The deployment strategy follows a natural progression from small voluntary groups (Stage~1) through organizations and institutions (Stage~2) to municipal and civic deployments (Stage~3) and larger civic use (Stage~4). Each stage informs the next. The lessons from early deployments shape the platform's development before it reaches larger scales.

If the platform is not good enough for its own contributors to use, it is not ready for anyone else.

\section{Governance of Votiverse Itself}

A platform for democratic governance should, eventually, govern itself democratically. The long-term aspiration is for Votiverse to be governed through its own mechanisms---a Votiverse instance managing the platform's roadmap, policies, and evolution. The platform is an open-source project by Diego Macrini, sponsored by Proximify Inc.\ (Ottawa, Canada), and operated through the Votiverse Foundation as an open initiative.

\section{Related Work}

Votiverse builds on a substantial body of prior work. \textbf{Liquid democracy:} The concept of revocable, transitive, topic-specific delegation has been explored theoretically \cite{tullock1967,miller1969,ford2002} and in software platforms such as LiquidFeedback \cite{behrens2014}. Theoretical and empirical analyses have examined liquid democracy from epistemic, equality, and formal perspectives \cite{blum2016,kling2015,paulin2020}. Votiverse extends this work by embedding liquid delegation as one configuration within a broader governance space, and by adding the prediction-tracking accountability layer that liquid democracy literature has not addressed.

\textbf{Swiss direct democracy:} Switzerland's system of regular referenda with mandatory voter booklets is the operational inspiration for Votiverse's voting events. Votiverse digitizes and extends the Swiss model with delegation, predictions, and community notes.

\textbf{DAO governance:} Decentralized autonomous organizations have implemented delegation mechanisms similar to liquid democracy. Votiverse draws on this experience but is not limited to blockchain-based identity or token-weighted voting.

\textbf{Participatory budgeting:} Municipal programs worldwide have demonstrated that citizens can meaningfully engage with specific policy decisions when given structured information and accessible tools.

\textbf{Prediction markets and forecasting:} The use of predictions to evaluate decision quality draws on the logic of forecasting tournaments and prediction markets, adapted to a governance context.

\textbf{Governance experimentation platforms:} VoteLab~\cite{kunz2023} is an open-source platform for experimenting with different voting input formats (majority, approval, range, Borda count) on smartphones, with a proof-of-concept studying voting outcome consistency. It addresses the experimentation gap for \emph{voting methods}; Votiverse addresses the gap for \emph{governance configurations}---compositions of delegation, deliberation, accountability, and awareness mechanisms.

\textbf{Grassroots digital democracy:} Shapiro~\cite{shapiro2024} proposes a grassroots architecture for local digital communities that federate into a global digital democracy, operating solely on members' smartphones without centralized servers. The architecture addresses infrastructure---consensus, social networking, currencies---while Votiverse addresses governance mechanisms within whatever infrastructure hosts it. The two approaches are complementary.

\textbf{Digital democracy as a research program:} Grossi et~al.~\cite{grossi2024} present a multi-methods research vision for digital democracy technology, co-authored by 29 researchers across computational social choice, political science, and AI. Their call for empirically and computationally informed development of digital democracy aligns with Votiverse's emphasis on configurable, measurable governance.

\section{Conclusion}

Democracy is a technology. Like all technologies, it can be improved. The systems we have inherited were designed for a world of limited communication, slow information flow, and geographically constrained communities. Votiverse does not claim to have solved democracy. It claims that the configuration space of democratic governance is far larger than what any single country currently explores, and that a platform enabling organizations to navigate that space---with accountability, transparency, and adaptability---is worth building.

A unifying design principle runs through every mechanism described in this paper: the platform must be \textbf{self-sustaining}. It cannot depend on external administrators for content curation, on external data sources for prediction verification, or on external authority for dispute resolution. Surveys generate the evidence. Community notes distribute the verification. Delegate candidacies subject leadership claims to scrutiny. Immutability makes all of it trustworthy. In a self-sustaining system, the act of participation is simultaneously an act of system maintenance---participants maintain the system not as a separate chore, but by using it.

\appendix
\section{Formal Model}
\label{app:formal}

This appendix provides a mathematical specification of the governance model.

\subsection{Participants and Issues}

Let $P = \{p_1, p_2, \ldots, p_n\}$ be the set of participants and $I = \{i_1, i_2, \ldots, i_m\}$ the set of issues to be decided. Each issue $i$ has at most one topic: $T(i) \in \mathcal{T} \cup \{\bot\}$, where $\mathcal{T}$ is a finite hierarchical topic taxonomy and $\bot$ denotes an unclassified issue.

\subsection{Delegations}

A delegation is a tuple $d = (p_s, p_t, \sigma)$ where $p_s \in P$ is the delegating participant (source), $p_t \in P$ is the delegate (target) with $p_t \neq p_s$, and $\sigma$ is the scope, which takes one of three forms: $\text{issue}(i)$ for a specific issue~$i$ (highest precedence), $\text{topic}(t)$ for $t \in \mathcal{T}$ covering issues classified under topic~$t$ or any of its descendants, or $\text{global}$ covering all issues (lowest precedence). All three scope types are transitive.

Let $D$ be the set of all active delegations. $D$ satisfies a \emph{uniqueness invariant}: for any participant $p_s$ and scope $\sigma$, there is at most one delegation $(p_s, \cdot, \sigma) \in D$. Creating a new delegation with the same source and scope replaces any existing one.

A delegation is \emph{active} for issue $i$ according to its scope: an issue-scoped delegation $\text{issue}(i)$ is active only for issue~$i$; a topic-scoped delegation $\text{topic}(t)$ is active if $T(i) \neq \bot$ and $T(i)$ is equal to or a descendant of~$t$; a global delegation is active for all issues.

When multiple delegations from the same source are active for the same issue, they necessarily differ in specificity (by the uniqueness invariant). Precedence is resolved by specificity: $\text{issue} > \text{child topic} > \text{parent topic} > \text{global}$.

\subsection{The Delegation Graph}

For a given issue $i$, define the \emph{effective delegation} of participant $p_s$ as the highest-precedence active delegation from $p_s$ for issue $i$, if any. Let $\text{eff}(p_s, i)$ denote the target of this delegation. The delegation graph is $G_i = (P, E_i)$ where:
\[
E_i = \{(p_s, \text{eff}(p_s, i)) \mid p_s \text{ has at least one active delegation for issue } i\}
\]
Each participant has at most one outgoing edge per issue (the highest-precedence active delegation). The graph $G_i$ is therefore a collection of directed trees (a forest), with delegates at the roots.

\subsection{Vote Resolution}

Let $V_i \subseteq P$ be the set of participants who cast a direct vote on issue $i$.

\textbf{Override Rule.} A direct vote always overrides a delegation. If $p_s \in V_i$, then the edge $(p_s, p_t) \in E_i$ is removed before weight computation.

\textbf{Transitive Weight.} After applying the override rule, let $G_i'$ be the resulting graph. The effective weight $w(p, i)$ of participant $p$ on issue $i$ is:
\[
w(p, i) = 1 + \sum_{q \in \text{sources}(p, G_i')} w(q, i)
\]
where $\text{sources}(p, G_i') = \{q \mid (q, p) \in E_i'\}$.

\textbf{Non-Participation.} A participant $p$ who neither votes directly nor is reachable from a voting participant through delegations has $w(p, i) = 0$.

\subsection{Delegation Cycles}

Cycles are resolved by treating them as mutual non-delegation. If participants form a cycle and none cast a direct vote, all are treated as abstaining. If any participant in the cycle votes directly, that breaks the cycle.

\subsection{Properties}

The model satisfies:
\begin{enumerate}
    \item \textbf{Sovereignty.} $\forall p \in P, \forall i \in I$: $p$ can cast a direct vote, setting $w(p, i) = 1$ independently of any delegation.
    \item \textbf{One person, one vote.} $\sum_{p \in V_i'} w(p, i) \leq |P|$, with equality when all participants either vote or are reached by a delegation chain ending in a voter.
    \item \textbf{Monotonicity.} Casting a direct vote never reduces a participant's influence.
    \item \textbf{Revocability.} Any delegation $d \in D$ can be removed at any time $t < t_{\text{close}}$, and the system recomputes all weights.
\end{enumerate}

\bibliographystyle{plainnat}

\end{document}